\documentclass[aps,prd,twocolumn,showpacs,superscriptaddress,letterpaper,floatfix]{revtex4}

\usepackage{graphicx}
\usepackage{longtable}
\usepackage[centertags]{amsmath}
\usepackage{color}
\usepackage[latin1]{inputenc}
\usepackage[english]{babel}
\usepackage{xspace}
\usepackage{verbatim}

\newcommand{\I}{\ensuremath{{\rm i}\,}}

\newcommand{\mc}{mode cleaner\xspace}

\newcommand{\lag}{LG$_{33}$\xspace}
\newcommand{\tem}{LG$_{00}$\xspace}
\begin{document}

\title[Experimental demonstration of higher-order Laguerre--Gauss mode interferometry]
{Experimental demonstration of higher-order Laguerre--Gauss mode interferometry}

\author{Paul Fulda}
\affiliation{School of Physics and Astronomy, University of
Birmingham, Edgbaston, Birmingham B15 2TT, UK}
\author{Keiko Kokeyama}
\affiliation{School of Physics and Astronomy, University of
Birmingham, Edgbaston, Birmingham B15 2TT, UK}
\author{Simon Chelkowski}
\affiliation{School of Physics and Astronomy, University of
Birmingham, Edgbaston, Birmingham B15 2TT, UK}
\author{Andreas Freise}
\affiliation{School of Physics and Astronomy, University of
Birmingham, Edgbaston, Birmingham B15 2TT, UK}

\date{\today}

\begin{abstract} 	
The compatibility of higher-order Laguerre--Gauss (LG) modes with
interferometric technologies commonly used in gravitational wave detectors is investigated.
In this paper we present the first experimental results concerning
the performance of the LG$_{33}$ mode in optical resonators. We show that the Pound-Drever-Hall
error signal for a LG$_{33}$ mode in a linear optical resonator is identical to
that of the more commonly used LG$_{00}$ mode, and demonstrate the feedback control
of the resonator with a LG$_{33}$ mode. We succeeded to increase the mode purity of a LG$_{33}$ mode generated using a 
spatial-light modulator from 51\% to 99\% upon transmission 
through a linear optical resonator. We further report the experimental verification that a triangular 
optical resonator does not transmit helical LG modes.
\end{abstract}

\maketitle

\section{Introduction}\label{sec:intro} 
The sensitivity of second generation gravitational wave detectors such as Advanced LIGO~\cite{Harry10} 
and Advanced Virgo~\cite{VIR-027A-09} will be limited by the thermal 
noises of the test masses~\cite{Rowan05}, therefore the gravitational wave community is involved in research 
into methods for reducing the effects of this noise. 
One proposed method for thermal noise reduction is to use so called `flat beams' in the main interferometer, in place of the currently
standard fundamental LG$_{00}$ beam~\cite{Vinet07}. This method
is currently investigated for possible upgrades of second generation detectors~\cite{LSCWP09v4}, as well as 
for third generation detectors~\cite{et_punturo2010}.
Flat beams
provide wider intensity distributions for the same optical losses 
and can therefore average better over the mirror surface
distortions caused by the thermal motions. A number of different flat beam shapes have been 
proposed, such as mesa beams~\cite{DAmbrosio03}, conical beams~\cite{Bondarescu08}
and higher-order LG beams~\cite{Mours06}. One advantage of higher-order LG beams over the other flat 
beams is their potential compatibility with the currently used spherical mirror surfaces. For this reason we 
find it worthwhile to further investigate the compatibility of higher-order LG beams with 
gravitational wave interferometer technology.
Some of us
recently showed that the potential detection rate of binary neutron star inspiral 
systems with the Advanced Virgo detector could be increased by a factor of 2.1 
if the LG$_{33}$ beam was used in place of the LG$_{00}$ beam~\cite{Chelkowski09}.
In addition to the thermal noise benefits, the wider intensity distributions of higher-order 
LG beams have been shown to reduce the magnitude of thermal aberrations of optics within the 
interferometers~\cite{vinet09}. This would reduce the extent to which thermal compensation 
systems are relied upon in future detectors to reach design sensitivity.
The improved thermal noise characteristics of higher-order LG beams over the fundamental beam also makes the 
technology a strong candidate for improving the sensitivity of optical clock experiments, which currently also approach 
the mirror thermal noise limit~\cite{webster08}.

An investigation using numerical simulations, into the sensing and control signals for length
and alignment with a LG$_{33}$ beam in advanced detectors
also yielded positive results, indicating that the LG$_{33}$ beam performed as well
if not better than the \tem beam in all of the examined criteria~\cite{Chelkowski09}.
We present the results of an experimental follow-on study into the interferometric performance of
higher-order LG beams, in order to assess how much of the potential for sensitivity
improvement is realizable in practice. The first crucial test for the interferometric
performance of \lag modes is their compatibility with \mc technology.
We show that the mode cleaner effect works equivalently for the \lag mode as for
the \tem mode in a linear optical resonator as depicted in
figure~\ref{fig:LGmode-lin-cav}. We also demonstrate the non-compatibility
of helical LG modes with three mirror mode cleaners.

\section{Performance of higher-order Laguerre--Gauss beams in mode cleaner cavities}\label{sec:theory}
Laguerre--Gauss modes represent a complete set of solutions to the paraxial wave equation,
and as such are well suited to modeling the eigenmodes of spherical optical
resonators~\cite{Siegman}. There is some lack of consensus in current literature
about the exact naming of LG mode functions. Much of the literature relating
to higher-order LG modes refers to modes with spiral phase fronts, which carry orbital
angular momenta $l\hbar$ per photon, where $l$ is the azimuthal mode 
index~\cite{Turnbull96, Courtial99, Kennedy02}. Equation~\ref{eqn:lghxamp} shows in cylindrical polar coordinates the 
normalized form of the complex amplitude of this mode set, which in the following will be
referred to as \emph{helical} LG modes. 
\begin{equation} 
\begin{split}
u_{p,l}^{\textrm{hel}}(r,\phi,z) =\frac{1}{w(z)}\sqrt{\frac{2p!}{\pi(|l|+p)!}}\quad e^{
\I\left(2p+|l|+1\right)\Psi(z)}\\
\times \,\left(\frac{\sqrt{2}r}{w(z)}\right)^{|l|}L^{(|l|)}_{p}\left(\frac{2r^2}{w(z)^2}\right)
e^{-\textrm{i}k\frac{r^2}{2q(z)}+\textrm{i}l\phi}\\
\end{split} 
\label{eqn:lghxamp}
\end{equation} 
where $p\geq 0$ is the radial mode index, $l$ is the azimuthal mode index, $\Psi$ is the Gouy phase, 
$w$ is the beam radius, $k$ is the wavenumber and $q$ is the complex
Gaussian beam parameter.
$L^{(|l|)}_{p}$ 
are the associated Laguerre polynomials.
An alternative form of LG modes with a sinusoidal amplitude dependence in
azimuthal angle can be used equally well.
The normalized form of the complex amplitude
of this mode set is shown in equation~\ref{eqn:lgcosamp};
we will refer to this
mode set as \emph{sinusoidal} LG modes. The symbols are as defined in equation \ref{eqn:lghxamp}, and $\delta$ is the Kronecker 
delta.
\begin{equation}
\begin{split}
u_{p,l}^{\textrm{sin}}(r,\phi,z) =\frac{2}{w(z)}\sqrt{\frac{2p!}{1+\delta_{0l}\pi(|l|+p)!}}\,
e^{\I\left(2p+|l|+1\right)\Psi(z)}\\
\times \left(\frac{\sqrt{2}r}{w(z)}\right)^{|l|}
 L^{(|l|)}_{p}\left(\frac{2r^2}{w(z)^2}\right)
e^{-\I k\frac{r^2}{2q(z)}}\left\{
\begin{array}{l}
\sin(l\phi)\\
\cos(l\phi)
\end{array}
\right\}.
\end{split} 
\label{eqn:lgcosamp}
\end{equation}


A complete set of
sinusoidal solutions consists of the functions as given in equation~\ref{eqn:lgcosamp}
using $\cos(l\phi)$ for $l\geq0$ and $\sin(l\phi)$ if $l<0$.
Higher-order LG modes of both sets offer improvements in thermal noise
for gravitational wave interferometers compared to the \tem mode, however the advantage is greater
for the helical modes than for the sinusoidal modes~\cite{vinet09}.

A number of different methods for generating higher-order LG
modes have been demonstrated~\cite{Kennedy02}. However, so far the optimization
of higher-order LG beam sources has largely been in a different direction to
that which is required by the gravitational wave detector community. For example
the use of LG beams in the cold atoms and optics fields often requires
high-speed manipulation of the beam parameters and positions, whereas 
the use of LG modes in high-precision interferometry depends on mode purity
and stability.
One of the leading candidate methods for the latter is the
use of diffractive optics, or \emph{phase plates} for conversion from a \tem mode to
a higher-order LG mode, due to their stability, as well as potentially high conversion
efficiency and output mode purity. Other conversion methods include using
computer generated holograms~\cite{Arlt98}, spatial light modulators~\cite{Matsumoto08}
and astigmatic mode converters~\cite{Courtial99}. However, none of these
mode conversion methods are perfect, and some light inevitably remains in unwanted
modes. An effectively  pure and stable higher-order LG mode
light source for gravitational wave interferometers can possibly be achieved with the implementation
of \emph{mode cleaner cavities}. 

In practice, mode cleaners take the form of medium- to 
high-finesse optical resonators which are
feedback controlled to remain on resonance for a chosen laser mode~\cite{Ruediger81}.
Mode cleaners are used in several locations in gravitational
wave interferometers~\cite{Abbott04}. So-called \emph{pre-mode cleaners} are used 
in the initial frequency stabilization chain of the laser. These typically employ
small, monolithic spacers in air. The beam then passes the \emph{input mode cleaners},
suspended optical cavities in vacuum whose main function is to filter beam geometry
fluctuations (also called beam-jitter noise). Modern laser interferometers also
include optical cavities in the main interferometer, which act as
additional mode cleaning cavities. Often a small in-vacuum output mode cleaner is then used
to filter the light leaving the interferometer before it reaches the photo detectors.
Mode cleaners can in principle be used to increase the spatial purity of any Gaussian mode.
Experimental verification of the compatibility of higher-order LG beams 
with mode cleaner technology is of paramount 
importance for determining the future prospects for LG beams in gravitational wave interferometers.

\begin{table}
\begin{center}
\begin{tabular}{|l|c|c|c|c|c|}
\hline
Mode cleaner  & Finesse & FSR & TEM$_{01}$ & Throughput \\
& & & suppression & \\
\hline
GEO MC1~\cite{gossler03} & 2700 & 37.48\,MHz & 1325 & 80\,\%\\
GEO MC2~\cite{gossler03} & 1900 & 37.12\,MHz & 937 & 72\,\%\\
Virgo IMC~\cite{Genin10} & 1181 & 1.044\,MHz & NA & 86.6\,\%\\
AdLIGO IMC~\cite{AdLIGOrefdesign} & 500 & 17.96\,MHz & NA & NA\,\\
\hline
Linear MC & 172 & 714\,MHz & 50.1 & 63\,\%\\
Triangular MC & 300 & 714\,MHz & 87.6 & 99\,\%\\
\hline
\end{tabular}
\caption{Input mode cleaner parameters for some gravitational wave 
detectors, as well as those used in this work. TEM$_{01}$ suppression factors and 
throughput percentages are given in terms of light power. The finesse and 
TEM$_{01}$ suppression factors of the mode cleaners used in this work 
were chosen to be lower than those of the large-scale mode cleaners.}
\label{tab:inputMC}
\end{center}
\end{table}

Currently a triangular arrangement is favored for the mode cleaners in gravitational
wave detectors as it allows to  
spatially separate the injected beam from the reflected beam, enabling a length control 
error signal to be measured in reflection without the need for polarizing optics. 
However, triangular cavities are not ideal for use with higher-order modes.

\begin{figure}[htb]
\begin{center}
\includegraphics[width=0.14\textwidth,keepaspectratio]{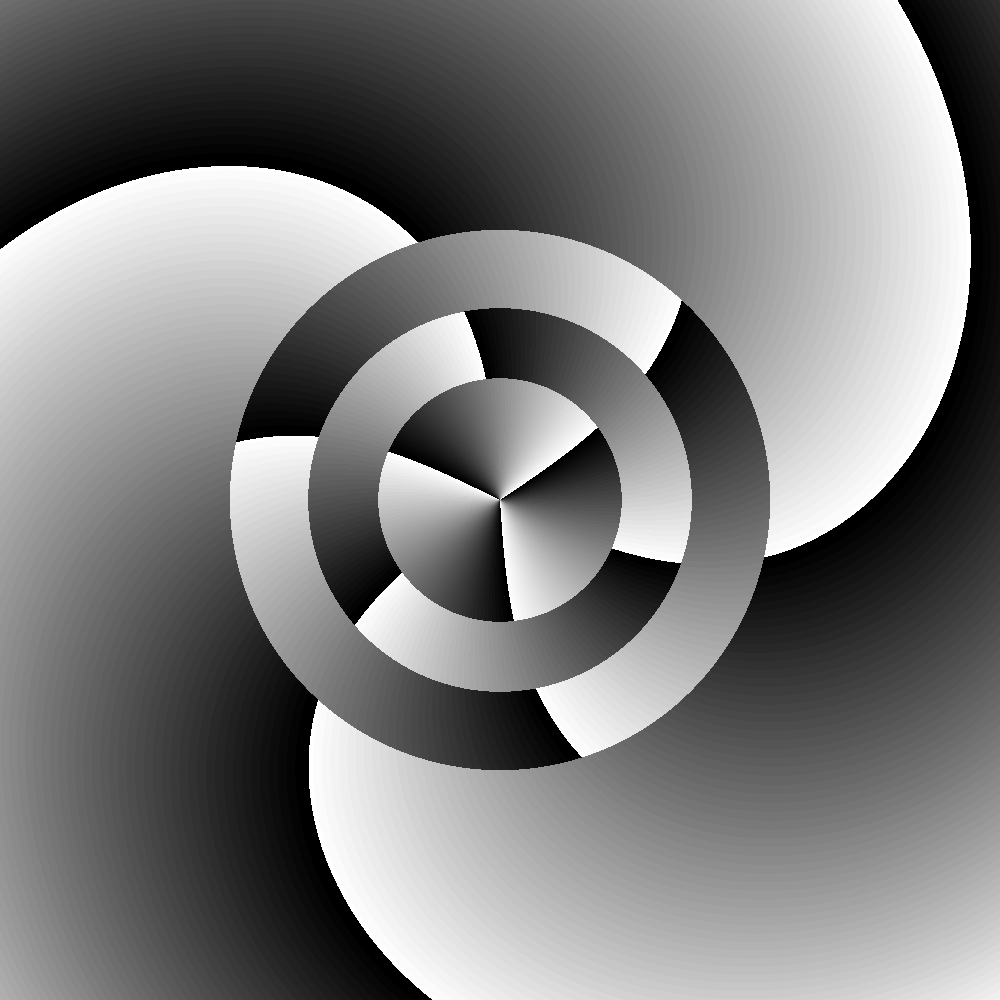}\hspace{2mm}
\includegraphics[width=0.14\textwidth,keepaspectratio]{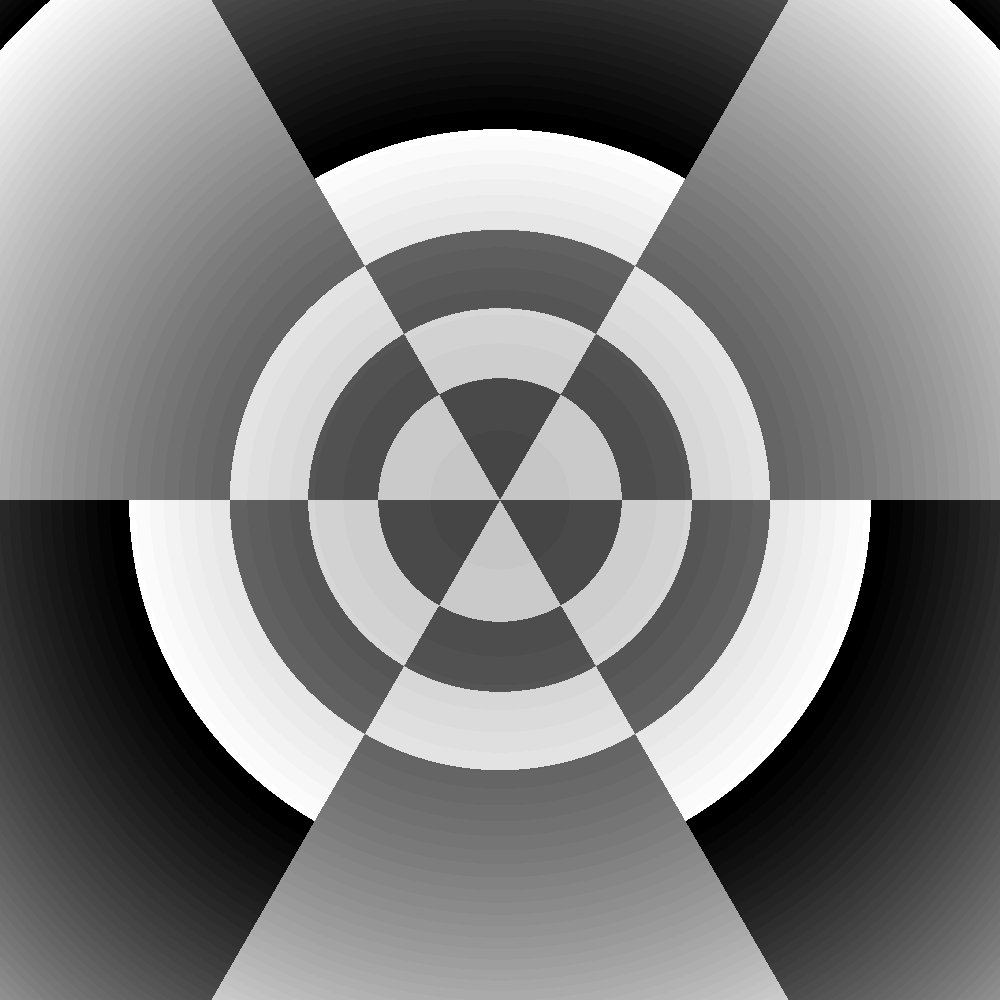}\hspace{2mm}
\includegraphics[width=0.14\textwidth,keepaspectratio]{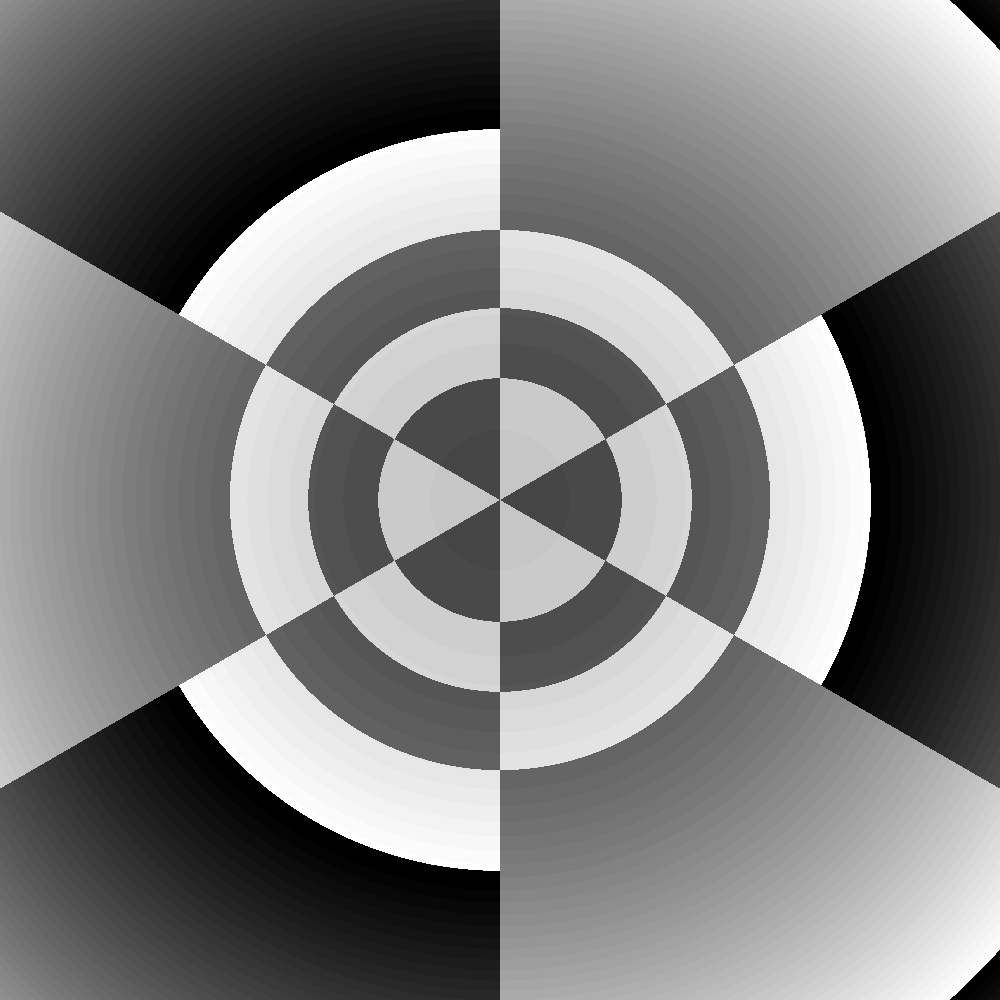}
\end{center}
\caption{Transverse phase distributions of the helical (left), vertically symmetric sinusoidal (center) and 
vertically anti-symmetric sinusoidal (right) \lag modes. The color represents the phase, in a range from 
0 (white) to 2$\pi$ (black).}
\label{fig:LG33-sincosphases}
\end{figure}

Triangular cavities behave differently from linear cavities in several ways. One important difference is that 
after one full round-trip in a triangular cavity any beam is mirrored about the vertical axis, which 
means that only light fields with symmetry about this axis can constructively interfere and be fully resonant. 
The intensity patterns of \lag modes are symmetric regarding a mirroring around the vertical axis. However, the phase
cross sections in general are not, as is shown in figure~\ref{fig:LG33-sincosphases}. Both types of sinusoidal modes
show the required symmetry about the vertical axis, but helical modes do not. The anti-symmetric sinusoidal mode will be resonant in a cavity
with an optical path length difference of $\lambda/2$ from one resonant for the symmetric mode. In other
words, a cavity tuned to be resonant for one type of sinusoidal mode will be anti-resonant for the respective other.
Furthermore any helical LG mode can be understood to be
a sum of two sinusoidal LG modes, by considering equations~\ref{eqn:lghxamp} and \ref{eqn:lgcosamp} and the identity 
$\exp(\I x)=\cos(x)+\I \sin(x)$. 
We thus expect that in the case of a helical LG input beam, the mode cleaner cavity can be tuned
to a length at which one of the constituent sinusoidal LG modes will be resonant and thus transmitted
while the other constituent sinusoidal LG mode will be exactly anti-resonant and
thus reflected, i.e. the helical LG beam would be decomposed into
the two constituent sinusoidal modes. This effect can be generalized to optical resonators with any number of 
mirrors. In all resonators with an \emph{even} number of mirrors the effect of mirroring the beam about the vertical axis 
is canceled out in one full round trip, so we expect these to transmit helical LG beams. In all 
resonators with an \emph{odd} number of mirrors the effect does not cancel out, so we do not expect them to transmit helical 
LG beams.

Another important difference between linear and triangular mode cleaner cavities is that the latter 
feature a spherically curved mirror which is probed by the circulating beam under an angle (not normal incidence). 
This results in a breaking of the symmetry about the azimuthal angle for the mode cleaner eigenmodes. This is not 
usually a problem for fundamental mode operation, since an astigmatic \tem mode is still an eigenmode of the cavity. 
Higher-order LG modes on the other hand are not eigenmodes of astigmatic cavities~\cite{astigm}. The mode shape of 
even sinusoidal LG beams degenerates upon transmission through a triangular mode cleaner as a result of the 
astigmatism. There are two possible solutions to this problem; to use linear cavities exclusively, or to design non-
astigmatic mode cleaner cavities with four or more mirrors. Some work has already been done to design non-astigmatic 
mode cleaner cavities for fundamental mode operation~\cite{skettrup05}, which should be investigated for use with 
higher-order LG modes. One possibility may be to implement aspherical mirrors to build a non-astigmatic mode cleaner 
for higher-order LG modes. It should be noted that using only linear cavities as mode cleaners incurs the additional 
complication of using polarizing optics to extract the control signals
in reflection.

As a result of these considerations, the main experimental setup described here makes use of a linear
mode cleaner cavity instead of a triangular cavity. We have however also experimentally verified
the non-transmission of helical modes through a triangular cavity (see section~\ref{sec:triangular}).
The finesse of the linear cavity was chosen to be low in comparison with some gravitational wave detector input mode 
cleaners, as shown in table \ref{tab:inputMC}. 
While higher finesse cavities 
can give a stronger suppression of misalignment modes, it is interesting to see
the large improvement that can already be gained through the use of a low-finesse mode cleaner.

\section{Laboratory demonstration}\label{sec:lab} 
In order to investigate the interferometric
performance of the LG$_{33}$ beam in a laboratory, it was necessary to
produce a reasonably pure example of such a beam. We used a computer-controlled
liquid-crystal-on-silicon spatial light modulator (SLM) for LG
beam preparation, as demonstrated in~\cite{Matsumoto08}, 
because of the availability and adaptability of such devices. 
We expect the SLM to be replaced
by a passive, etched phaseplate in eventual implementations of \lag modes in 
gravitational wave detectors.

The report on the laboratory investigation is comprised of two parts; the
first concerning the performance of the sinusoidal and helical \lag beams in a linear
mode cleaner, and the second concerning specifically the performance of the
helical \lag beam in a triangular mode cleaner.

\subsection{LG mode performance in a linear mode cleaner}\label{subsec:lin} 
\begin{figure}[htb]
\begin{center}
\includegraphics[width=0.48\textwidth,keepaspectratio]{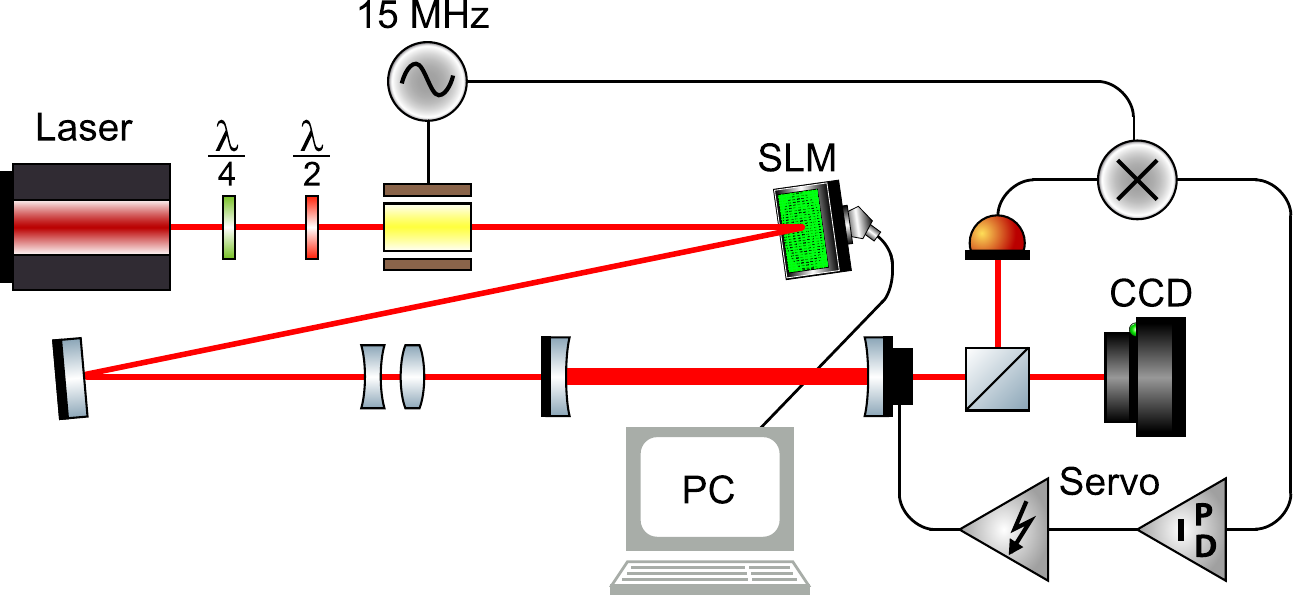}
\end{center}
\caption{The experimental setup for mode cleaning a SLM generated higher-order LG beam. 
The \tem input beam is converted to a higher-order LG beam
by the SLM. The resulting beam is passed through a mode-matching telescope into the linear cavity. 
The transmitted light is used to generate an error signal which is fed back to the PZT attached to the 
curved end mirror to control the length of the cavity. The transmitted beam is simultaneously imaged 
on the CCD camera.}
\label{fig:LGmode-lin-cav}
\end{figure}

The experimental
setup for the investigation into the performance of the \lag mode in
a linear mode cleaner is shown in figure~\ref{fig:LGmode-lin-cav}. The 1064\,nm
laser light is passed through quarter and half wave plates to set the polarization
vector to the optimum orientation for use with the SLM. An electro-optic modulator
(EOM) is used to imprint a 15\,MHz modulation on the light to enable length
control of the mode cleaner with the Pound-Drever-Hall (PDH) method~\cite{Black00}.
The light is then reflected from the modulating surface of the SLM, where
the phase characteristics of the desired LG mode are imprinted on the beam. 
The resulting beam is then passed through a telescope to match the
beam to the mode cleaner eigenmode.

The light transmitted through the mode cleaner is passed through a beam splitter, and
analyzed at the two ports with a CCD camera and a photodiode respectively. 
The signal from the photodiode is mixed down with the 15\,MHz signal to generate the
PDH error signal, which is then fed back to a Piezo-electric transducer (PZT)
attached to the mode cleaner end mirror, via a servo and high-voltage amplifier. 
In this way the length of the mode cleaner cavity can be controlled to maintain
the resonance condition for a given mode order. 
In typical implementations of mode cleaners in gravitational wave interferometers,
the error signal is taken in reflection. For this work, however, the mode cleaner cavity was of
a low enough finesse to allow the error signal to be taken in transmission.

\begin{figure}[htb]
\begin{center}
\includegraphics[width=0.38\textwidth,keepaspectratio]{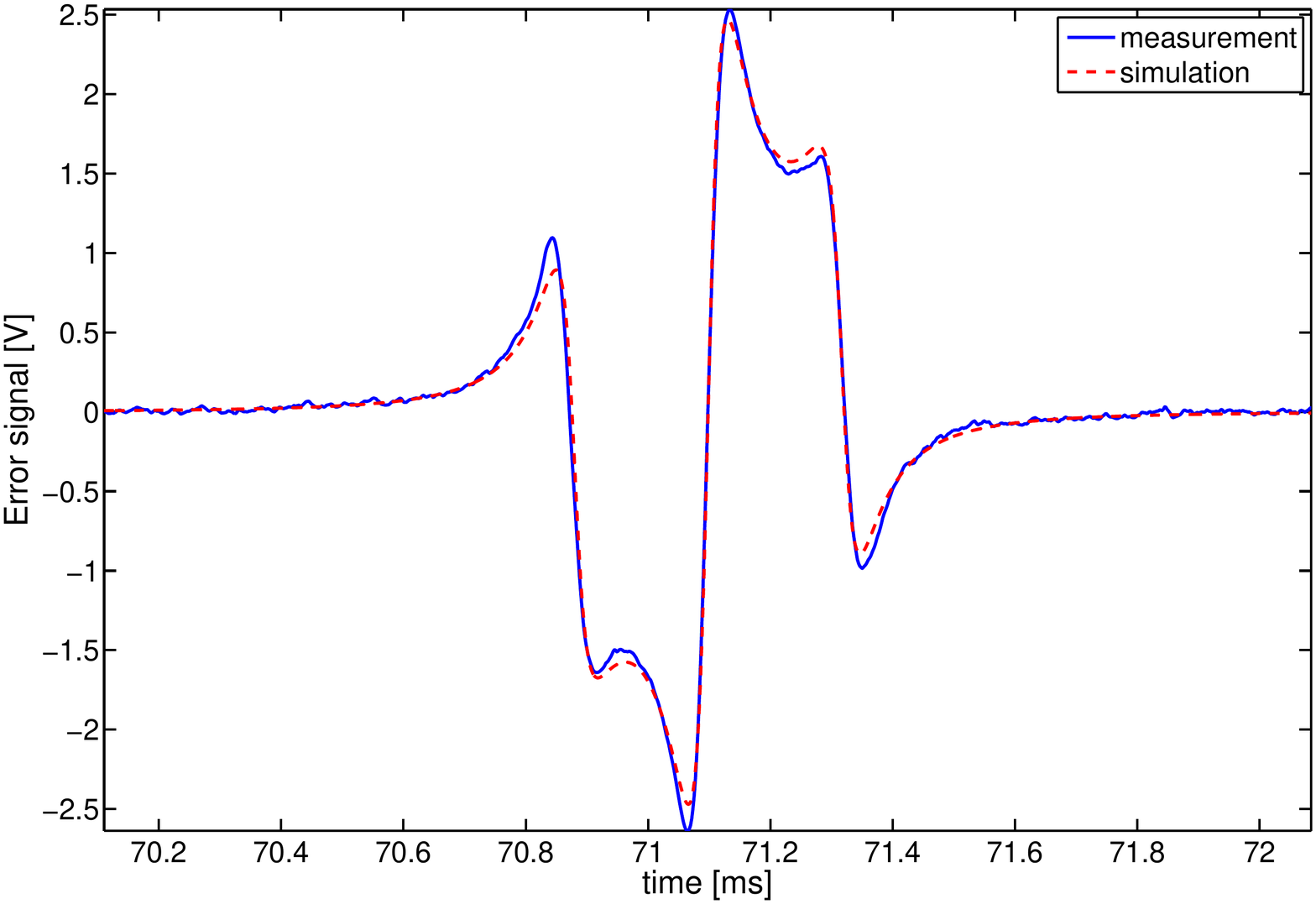}
\caption{The blue trace shows the PDH error signal from the linear cavity, set up as shown in figure 
\ref{fig:LGmode-lin-cav}, with a sinusoidal \lag input beam. The red dashed trace shows the PDH 
error signal for the same optical setup as simulated in the frequency domain simulation software 
Finesse \cite{Finesse}. While there are small discrepancies between the two traces, the primary 
features are identical, as predicted in \cite{Chelkowski09}.}
\label{fig:sinusoidalPDH}
\end{center}
\end{figure}

The PDH error signal for a sinusoidal \lag input beam is shown in figure~\ref{fig:sinusoidalPDH}.
This signal was recorded from the output of the mixer while scanning
over the \lag resonance of the linear mode cleaner. 
The error signal is equivalent to that generated when the input
beam is a \tem beam, confirming the result in~\cite{Chelkowski09}, and thus
allowed a robust feedback control of the cavity length.
This is a significant result, as the PDH control loop method is a fundamental
technique in the operation of gravitational wave interferometers.

\begin{figure}[htb] 
\begin{center} 
\includegraphics[width=0.21\textwidth,keepaspectratio]{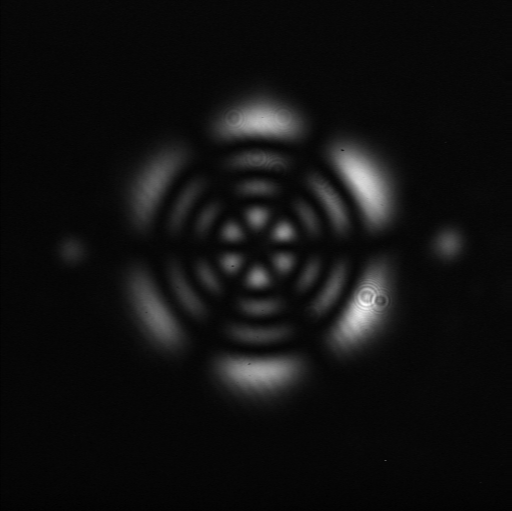}\hspace{0.05cm}
\includegraphics[width=0.21\textwidth,keepaspectratio]{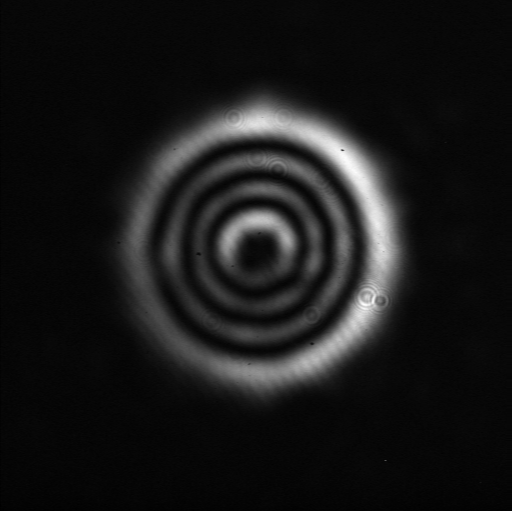}\vspace{0.1cm}
\includegraphics[width=0.21\textwidth,keepaspectratio]{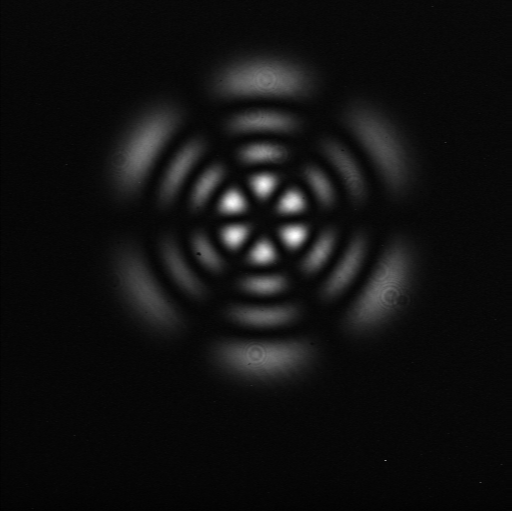}\hspace{0.05cm}
\includegraphics[width=0.21\textwidth,keepaspectratio]{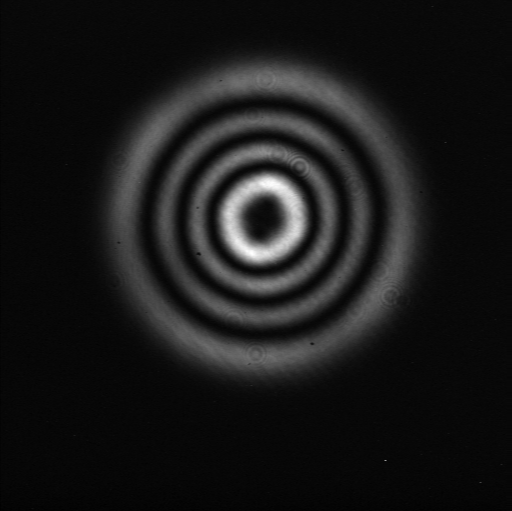}\vspace{0.1cm}
\caption{The measured intensity patterns of the sinusoidal (left column) and helical (right
column) \lag beams before (upper row) and after (lower row) transmission through
the linear mode cleaner. The increase in mode purity upon transmission is already
evident in the increased symmetry. The remaining asymmetry apparently is a result of the inaccuracy in 
the manual alignment of the input beam to the mode cleaner. This effect is the same for both images
but more visually apparent in the case of the helical mode.} 
\label{fig:modepics}
\end{center} 
\end{figure} 

\begin{figure}[htb]
\begin{center}
\includegraphics[scale=0.215,viewport=162 0 530 300,clip]{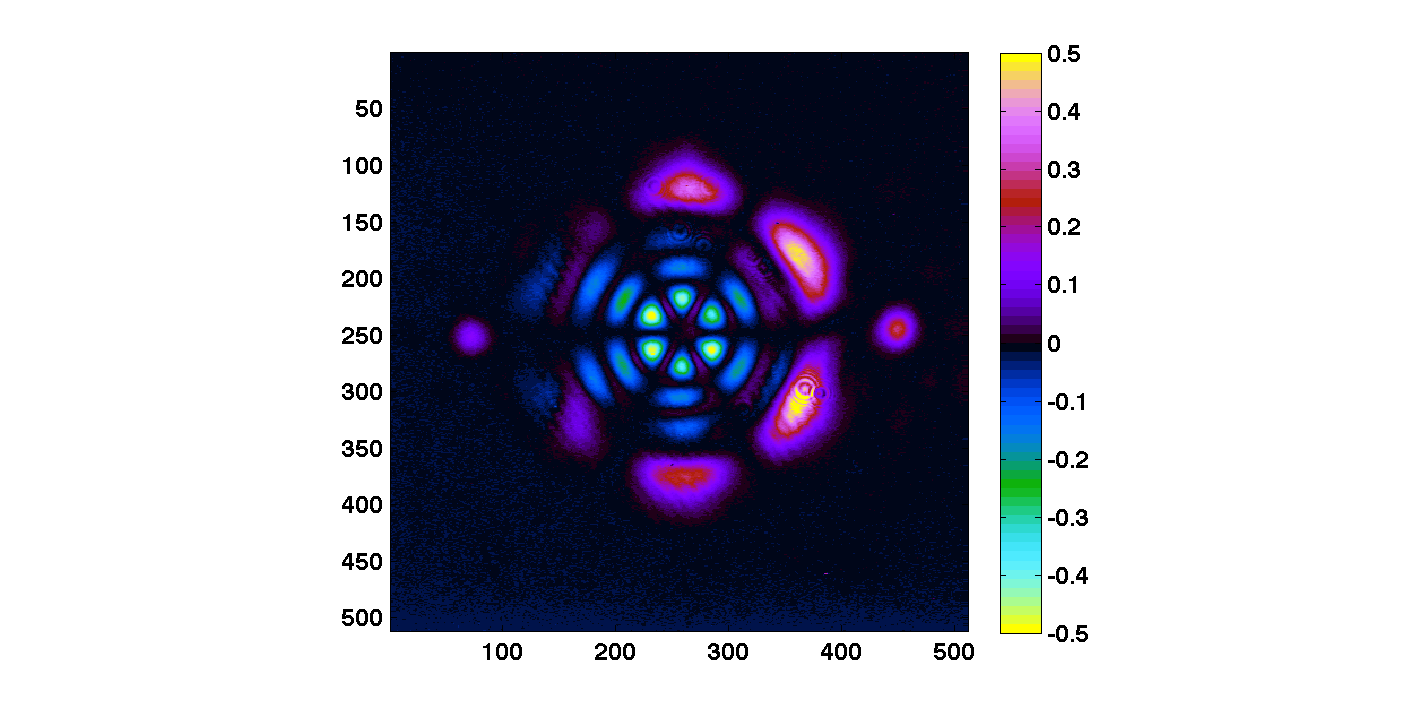}
\includegraphics[scale=0.215,viewport=163 0 530 300,clip]{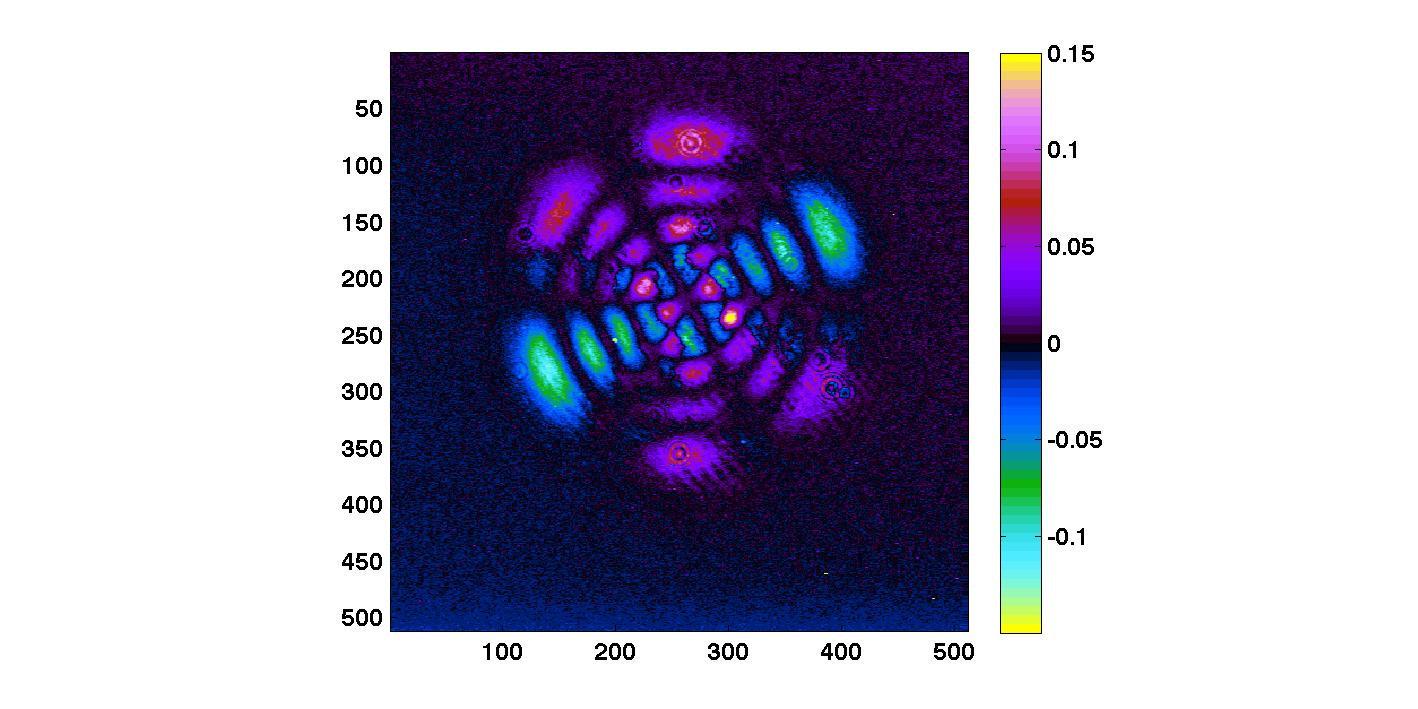}
\includegraphics[scale=0.215,viewport=165 0 530 300,clip]{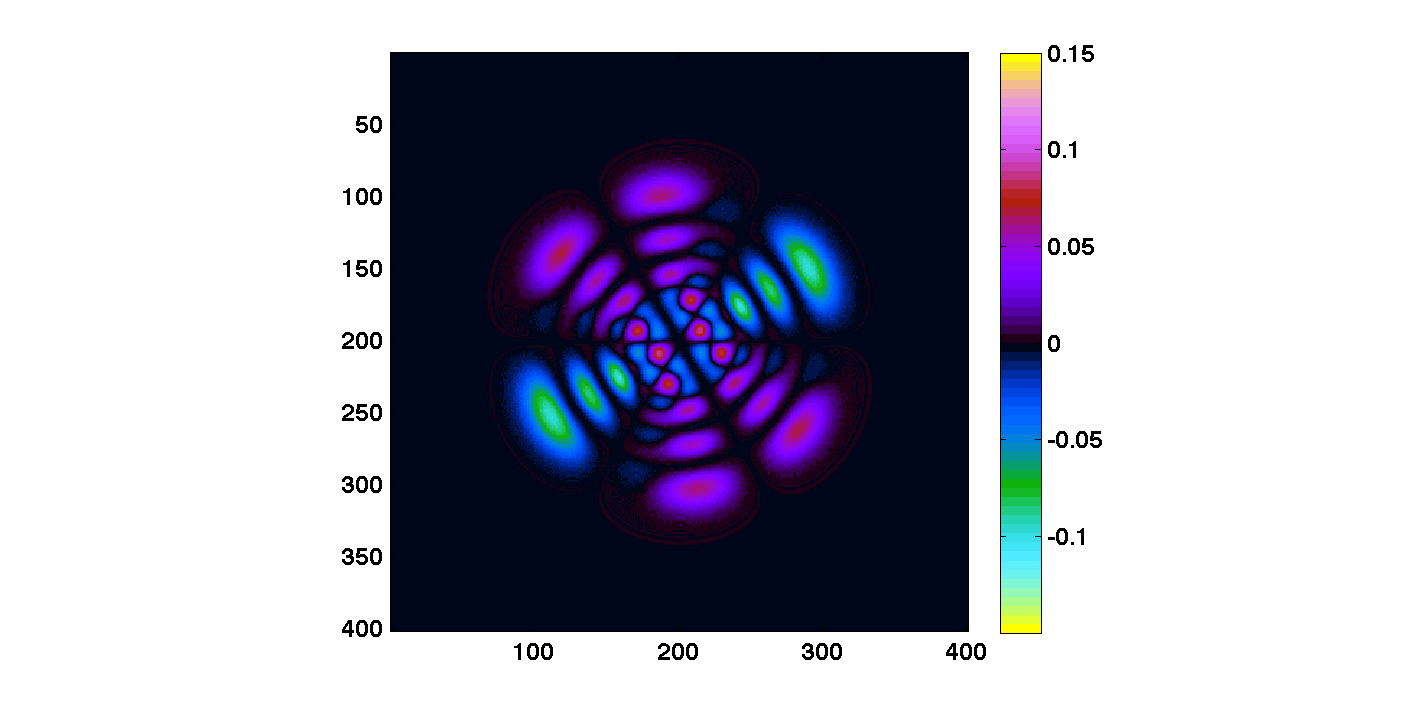}
\caption{Residuals from best fits between intensity patterns and a theoretically ideal sinusoidal 
\lag intensity pattern. From left to right: the residual for the measured input \lag beam, the residual for the measured 
output \lag beam, the residual for an output \lag pattern generated with a numerical model including a
misalignment of the input beam to the cavity.}
\label{fig:residuals}
\end{center}
\end{figure}

The CCD camera was used to record intensity images of the transmitted beams while the
mode cleaner was controlled to be resonant for the \lag mode.
Figure~\ref{fig:modepics} shows the input and output beam intensity patterns for
both helical and sinusoidal \lag beams. The images indicate that the 
output modes are more symmetrical, and have a higher intensity
in the innermost bright radial fringe relative to the others; a feature that
is characteristic of \lag modes. 
The typical method for measuring the output mode purity would be to pass the 
output beam through another cavity and observe the 
magnitudes of different mode order resonances~\cite{KSWD07}.
This method in its original form is not ideal for the work described in this paper however,
since in this case the performance of the mode in a cavity is itself being investigated.
Instead,
we have been able to estimate the mode content based on the intensity pattern alone
using numerical simulations. The light transmitted through the cavity can be described well
using eigenmodes of said cavity. We were able to construct a numerical model representing 
the measured intensity pattern as a sum of LG eigenmodes using the following steps. 
We initially performed non-linear fits of the
\lag eigenmode to the measured intensity pattern: the model function
was a pure LG$_{33}$ beam and the beam center as
well as the beam size have been adjusted by the fitting
algorithm. This step effectively produces a calibration of the
position and pixel size of the CCD sensor. Once the beam position
and size in the coordinate system of the sensor are known, the
LG$_{33}$ model can be subtracted from the measured data.
Figure~\ref{fig:residuals} shows the residuals of this step: the left panel
shows the residual between the input sinusoidal \lag beam intensity pattern shown
in figure~\ref{fig:modepics} and a theoretical \lag mode. The central
panel shows the equivalent residual for the output sinusoidal \lag beam intensity
pattern. It can be seen that the scale of the residuals is less
for the output \lag beam than for the input beam. The residual of the 
transmitted beam also indicates that the mode deformation is dominated by a
misalignment of the injected beam into the mode cleaner. 
Using the interferometer simulation Finesse~\cite{Finesse} a model
of the setup was used to search the alignment parameters space.
It was possible to
create a beam pattern like the measured pattern 
when the model included
an input beam misaligned by
$\alpha_x=-100\,\mu$rad in the horizontal plane, and $\alpha_y=60\,\mu$rad in the
vertical plane. The residual pattern between the intensity pattern calculated
with the Finesse simulation, and an ideal \lag mode is shown
in the right hand panel of figure~\ref{fig:residuals}.
Based on the Finesse result we were able to produce a very good numerical model of the transmitted
field amplitude and estimate the mode content by separately evaluating the overlap integrals between the 
complex field amplitude of the model and the field amplitudes of all LG eigenmodes. The results for the
sinusoidal beam are shown in table~\ref{tab:lg33decompose}. Our model predicts that
99\% of the light power is in the \lag mode
and most of the remaining light power is distributed in other modes of the order 9. 
A similar analysis for the helical mode gave effectively the same results for the mode
purity.
	
\begin{table}[htb]
\begin{center}
\begin{tabular}{|l|c|c|c|c|c|c|}
\hline
$u^{\rm sin}_{l p}$ mode & 3, 3 & 4, -1 & 2, -5& 4, 1& 2, 5& other\\
\hline
power &  99\% & 0.4\% & 0.3\% & 0.1\% & 0.1\% &  $<10$\,ppm \\
\hline
\end{tabular}
\caption{Mode decomposition of the numerical model of the sinusoidal \lag beam transmitted through 
the linear mode cleaner, under an input beam misalignment of -100\,$\mu$rad in the 
horizontal axis, and 60\,$\mu$rad in the vertical axis. The majority of the 
beam power is in the desired sinusoidal \lag mode, with the rest almost entirely concentrated 
in other modes of order 9.}\label{tab:lg33decompose}
\end{center}
\end{table}

Since the transmitted beam was to 99\% in a single mode we were able to make an accurate
estimate of the input mode purity by
comparing the throughput of the \lag modes to that of the \tem mode.
Once the intrinsic optical losses
of the mode cleaner cavity were taken into account, we estimated the input
mode purity to be 51\% for the sinusoidal \lag beam, and 66\% for
the helical \lag beam. It should be noted that examples of higher-order LG modes with 
mode purities likely to be above 70\% have been created previously directly with SLMs 
using a more thoroughly optimized conversion
procedure, for example in~\cite{Matsumoto08}, although in this case
the authors refrain from quoting an experimentally measured purity. However, this is the
first time a purity improvement of a Laguerre-Gauss mode using an optical resonator
to an estimated 99\% has been {reported in the scientific literature}. The demonstrated mode 
purity is limited in first order by the manual 
alignment of the input beam and can very likely be improved using a standard automatic 
alignment system.

\subsection{Helical LG mode performance in a triangular mode cleaner} \label{sec:triangular}

\begin{figure}[htb]
\begin{center} 
\includegraphics[width=0.48\textwidth,keepaspectratio]{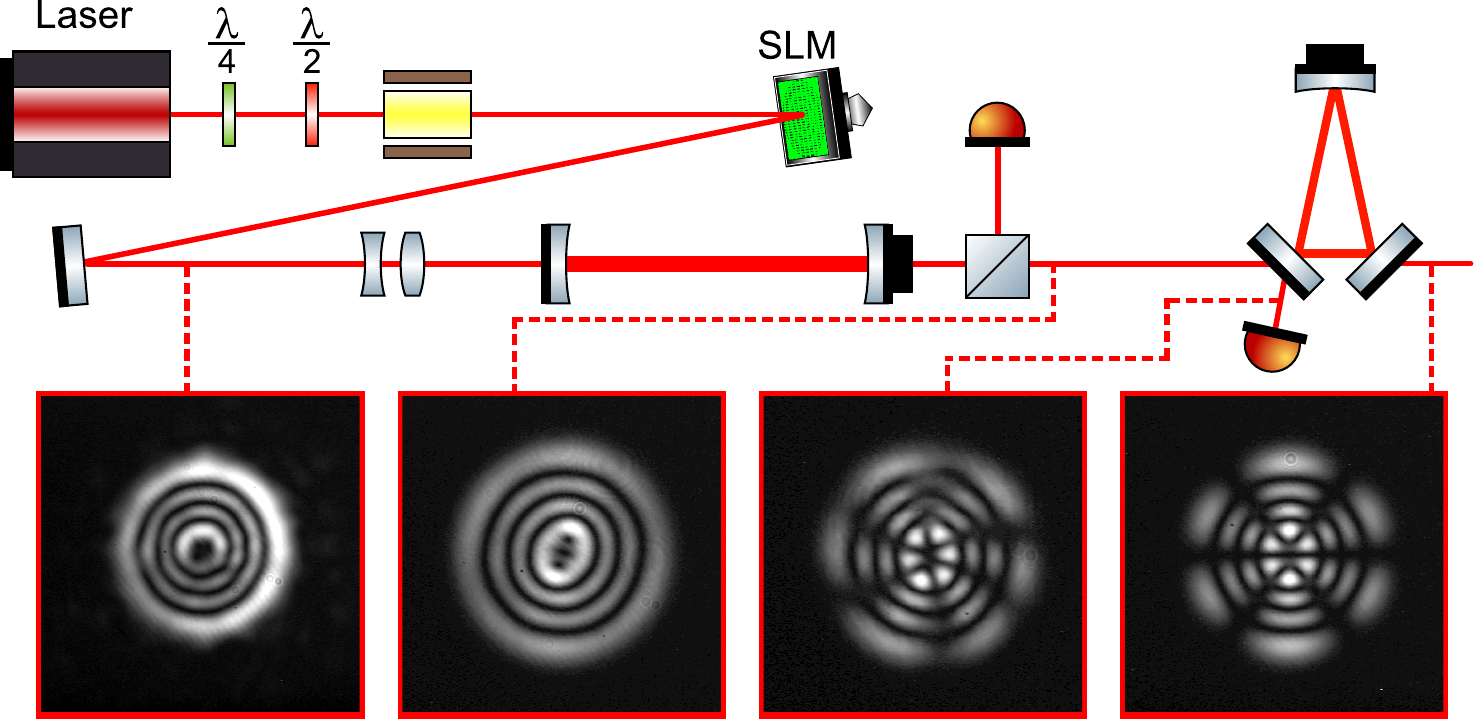}
\end{center} 
\caption{The experimental setup for transmitting a helical \lag mode through a triangular mode cleaner,
showing the intensity pattern of the beams at various locations in the setup (note that the images
shown here are contrast enhanced to show the pattern more clearly). From left
to right: helical \lag after the SLM, helical \lag after transmission through the linear
mode cleaner, beam reflected from the triangular cavity and beam transmitted through the triangular cavity.}
\label{fig:LG33hx-refltrans}
\end{figure}

In order to test the effect described in section \ref{sec:theory}, whereby we expected helical LG beams to be decomposed into 
their consitutuent sinusoidal LG modes upon interaction with a triangular mode cleaner,
we placed a cavity of the standard triangular pre-mode
cleaner design~\cite{WUGBKSS98} after the linear mode cleaner as shown in figure~\ref{fig:LG33hx-refltrans}. The triangular 
mode cleaner was scanned with the
sinusoidal \lag beam input and the helical \lag beam input successively. 
As expected, extra resonances at half a free-spectral range were observed 
when the input was changed from sinusoidal \lag to helical \lag. 
The triangular mode cleaner was then feedback controlled in similar fashion to the linear mode cleaner,
with the PDH error signal being this time obtained from the light reflected from the cavity input mirror.
Figure~\ref{fig:LG33hx-refltrans} shows images of the
input, transmitted and reflected beams at one of these resonances (the beam after the linear cavity
was of slightly lower quality than that shown in figure~\ref{fig:modepics} as less time was spent on the 
alignment optimization for this experiment).
We observed that the beam transmitted through the triangular cavity was nearly a vertically symmetric sinusoidal
\lag mode. The stronger vertical central section compared to the one shown in figure~\ref{fig:modepics} is caused by 
astigmatism due to the curved end mirror~\cite{astigm}.
The reflected beam was a superposition of all the modes rejected by the mode cleaner, and was therefore of lower 
mode purity than the transmitted mode. However, the vertically anti-symmetric \lag mode can be seen to be the dominant 
mode present in the reflected light. 
The measurement was repeated for the alternative resonance point, where as expected the dominant
type of the transmitted and reflected mode was reversed.
The helical input beam is decomposed into the
constituent sinusoidal modes upon interaction with the triangular mode cleaner, as predicted. We therefore
conclude that in order for helical \lag modes to be compatible with gravitational
wave interferometers, the mode cleaners used must be linear, or at least be comprised of an even
number of mirrors. If helical Laguerre-Gauss beams are to be implemented in second generation detector upgrades or 
third generation detectors, this result shows that at
the very least a redesign of the mode cleaners from the first and second generation detectors will be necessary, since 
these are currently triangular cavities. As mode cleaners are present in gravitational wave interferometers in several 
places, this constitutes a significant consideration for the overall optical design of the detectors for which LG mode technology 
is considered.
\section{Conclusion}\label{sec:conclusion}
Research into the interferometric performance of LG modes is important for the gravitational wave
community, as LG modes offer a thermal noise advantage over the fundamental \tem mode that can
improve the sensitivity achievable by future detectors~\cite{Mours06}. Simulations have already shown promising
results for the interferometric performance of LG modes~\cite{Chelkowski09}. This article
provides experimental verification for some of these predictions:
We have demonstrated the generation of a PDH error
signal from a linear mode cleaner injected with both helical and sinusoidal \lag modes equivalent to 
the error signal obtained with a \tem mode. We used this error signal to successfully demonstrate longitudinal 
control of the linear mode cleaner cavity at resonance for the \lag mode; a vital technique for the operation
of gravitational wave interferometers with \lag modes. We also showed an increase in the purity of a sinusoidal 
\lag mode from 51\% to 99\% upon transmission through a linear mode cleaner, demonstrating that very high-purity
\lag mode light sources can be produced in this way. 
Furthermore, we have demonstrated the decomposition of a helical \lag mode 
into the constituent sinusoidal \lag modes with a triangular mode cleaner; a result which has a strong impact on the 
choice of the optical design of future detectors.

The prospects for LG modes in gravitational wave detectors remain intact following the investigation described in this paper.
In the future we will expand this work to use LG modes in systems that combine Michelson interferometers and resonant cavities, and also
include alignment control systems. The
requirements for mirror surfaces for high finesse systems with LG modes also needs further investigation
to move scrutiny of the interferometric performance of LG modes to the next level, towards a possible
implementation in future gravitational wave detectors. 

\section{Acknowledgments} 
We would like to thank M. Padgett and the optics group from
Glasgow University for their assistance and advice. We would also like to thank
J. Nelson from the Institute for Gravitational Research for his support with the SLM. 
This work has been supported by the Science
and Technology Facilities Council and the European Commission 
(FP7 Grant Agreement 211743). This document has been assigned the LIGO 
Laboratory document number LIGO-P1000040
\bibliographystyle{apsrev}

\begin{thebibliography}{18}
\expandafter\ifx\csname natexlab\endcsname\relax\def\natexlab#1{#1}\fi
\expandafter\ifx\csname bibnamefont\endcsname\relax
  \def\bibnamefont#1{#1}\fi
\expandafter\ifx\csname bibfnamefont\endcsname\relax
  \def\bibfnamefont#1{#1}\fi
\expandafter\ifx\csname citenamefont\endcsname\relax
  \def\citenamefont#1{#1}\fi
\expandafter\ifx\csname url\endcsname\relax
  \def\url#1{\texttt{#1}}\fi
\expandafter\ifx\csname urlprefix\endcsname\relax\def\urlprefix{URL }\fi
\providecommand{\bibinfo}[2]{#2}
\providecommand{\eprint}[2][]{\url{#2}}

\bibitem[{\citenamefont{Harry and the LIGO
  Scientific~Collaboration}(2010)}]{Harry10}
\bibinfo{author}{\bibfnamefont{G.~M.} \bibnamefont{Harry}} \bibnamefont{and}
  \bibinfo{author}{\bibnamefont{the LIGO Scientific~Collaboration}},
  \bibinfo{journal}{Classical and Quantum Gravity}
  \textbf{\bibinfo{volume}{27}}, \bibinfo{pages}{084006}
  (\bibinfo{year}{2010}),
  \urlprefix\url{http://stacks.iop.org/0264-9381/27/i=8/a=084006}.

\bibitem[{\citenamefont{{The Virgo Collaboration}}(2009)}]{VIR-027A-09}
\bibinfo{author}{\bibnamefont{{The Virgo Collaboration}}}, \bibinfo{type}{Tech.
  Rep.}, \bibinfo{number}{VIR-027A-09}, \bibinfo{institution}{Virgo} (\bibinfo{year}{2009}).
  
  \bibitem[{\citenamefont{Rowan et~al.}(2005)\citenamefont{Rowan, Hough, and
  Crooks}}]{Rowan05}
\bibinfo{author}{\bibfnamefont{S.}~\bibnamefont{Rowan}},
  \bibinfo{author}{\bibfnamefont{J.}~\bibnamefont{Hough}}, \bibnamefont{and}
  \bibinfo{author}{\bibfnamefont{D.}~\bibnamefont{Crooks}},
  \bibinfo{journal}{Physics Letters A} \textbf{\bibinfo{volume}{347}},
  \bibinfo{pages}{25 } (\bibinfo{year}{2005}), ISSN \bibinfo{issn}{0375-9601},
  \urlprefix\url{http://www.sciencedirect.com/science/article/B6TVM-4GH43KY-5/%
2/6d03d4edb0af65b2c524289e237b5dfb}.

    \bibitem[{\citenamefont{Vinet}(2007)}]{Vinet07}
\bibinfo{author}{\bibfnamefont{J.-Y.} \bibnamefont{Vinet}},
  \bibinfo{journal}{Classical and Quantum Gravity}
  \textbf{\bibinfo{volume}{24}}, \bibinfo{pages}{3897} (\bibinfo{year}{2007}).
  
\bibitem[{\citenamefont{LIGO Scientific Collaboration}(2009)}]{LSCWP09v4}
\bibinfo{author}{\bibnamefont{{The LIGO Scientific Collaboration}}}, \bibinfo{type}{Tech.
  Rep.}, \bibinfo{number}{LIGO-T0900276-v4}, \bibinfo{institution}{LSC} (\bibinfo{year}{2009}).
  
\bibitem[{\citenamefont{Punturo et~al.}(2010)\citenamefont{Punturo, Abernathy,
  Acernese, Allen, Andersson, Arun, Barone, Barr, Barsuglia, Beker
  et~al.}}]{et_punturo2010}
\bibinfo{author}{\bibfnamefont{M.}~\bibnamefont{Punturo}},
  \bibinfo{author}{\bibfnamefont{M.}~\bibnamefont{Abernathy}},
  \bibinfo{author}{\bibfnamefont{F.}~\bibnamefont{Acernese}},
  \bibinfo{author}{\bibfnamefont{B.}~\bibnamefont{Allen}},
  \bibinfo{author}{\bibfnamefont{N.}~\bibnamefont{Andersson}},
  \bibinfo{author}{\bibfnamefont{K.}~\bibnamefont{Arun}},
  \bibinfo{author}{\bibfnamefont{F.}~\bibnamefont{Barone}},
  \bibinfo{author}{\bibfnamefont{B.}~\bibnamefont{Barr}},
  \bibinfo{author}{\bibfnamefont{M.}~\bibnamefont{Barsuglia}},
  \bibinfo{author}{\bibfnamefont{M.}~\bibnamefont{Beker}},
  \bibnamefont{et~al.}, \bibinfo{journal}{Classical and Quantum Gravity}
  \textbf{\bibinfo{volume}{27}}, \bibinfo{pages}{084007}
  (\bibinfo{year}{2010}),
  \urlprefix\url{http://stacks.iop.org/0264-9381/27/i=8/a=084007}.
  
\bibitem[{\citenamefont{{D'Ambrosio}}(2003)}]{DAmbrosio03}
\bibinfo{author}{\bibfnamefont{E.}~\bibnamefont{{D'Ambrosio}}},
  \bibinfo{journal}{Physical Review D} \textbf{\bibinfo{volume}{67}},
  \bibinfo{pages}{102004} (\bibinfo{year}{2003}).

\bibitem[{\citenamefont{Bondarescu et~al.}(2008)\citenamefont{Bondarescu,
  Kogan, and Chen}}]{Bondarescu08}
\bibinfo{author}{\bibfnamefont{M.}~\bibnamefont{Bondarescu}},
  \bibinfo{author}{\bibfnamefont{O.}~\bibnamefont{Kogan}}, \bibnamefont{and}
  \bibinfo{author}{\bibfnamefont{Y.}~\bibnamefont{Chen}},
  \bibinfo{journal}{Physical Review D (Particles, Fields, Gravitation, and
  Cosmology)} \textbf{\bibinfo{volume}{78}}, \bibinfo{eid}{082002}
  (\bibinfo{year}{2008}),
  \urlprefix\url{http://link.aps.org/doi/10.1103/PhysRevD.78.082002}.

\bibitem[{\citenamefont{{Mours} et~al.}(2006)\citenamefont{{Mours},
  {Tournefier}, and {Vinet}}}]{Mours06}
\bibinfo{author}{\bibfnamefont{B.}~\bibnamefont{{Mours}}},
  \bibinfo{author}{\bibfnamefont{E.}~\bibnamefont{{Tournefier}}},
  \bibnamefont{and} \bibinfo{author}{\bibfnamefont{J.-Y.}
  \bibnamefont{{Vinet}}}, \bibinfo{journal}{Classical and Quantum Gravity}
  \textbf{\bibinfo{volume}{23}}, \bibinfo{pages}{5777} (\bibinfo{year}{2006}).

\bibitem[{\citenamefont{Chelkowski et~al.}(2009)\citenamefont{Chelkowski, Hild,
  and Freise}}]{Chelkowski09}
\bibinfo{author}{\bibfnamefont{S.}~\bibnamefont{Chelkowski}},
  \bibinfo{author}{\bibfnamefont{S.}~\bibnamefont{Hild}}, \bibnamefont{and}
  \bibinfo{author}{\bibfnamefont{A.}~\bibnamefont{Freise}},
  \bibinfo{journal}{Physical Review D (Particles, Fields, Gravitation, and
  Cosmology)} \textbf{\bibinfo{volume}{79}}, \bibinfo{eid}{122002}
  (pages~\bibinfo{numpages}{11}) (\bibinfo{year}{2009}),
  \urlprefix\url{http://link.aps.org/abstract/PRD/v79/e122002}.

\bibitem[{\citenamefont{Vinet}(2009)}]{vinet09}
\bibinfo{author}{\bibfnamefont{J.-Y.} \bibnamefont{Vinet}},
  \bibinfo{journal}{Living Reviews in Relativity} \textbf{\bibinfo{volume}{12}}
  (\bibinfo{year}{2009}),
  \urlprefix\url{http://www.livingreviews.org/lrr-2009-5}.

\bibitem[{\citenamefont{{Webster} et~al.}(2008)\citenamefont{{Webster},
  {Oxborrow}, {Pugla}, {Millo}, and {Gill}}}]{webster08}
\bibinfo{author}{\bibfnamefont{S.~A.} \bibnamefont{{Webster}}},
  \bibinfo{author}{\bibfnamefont{M.}~\bibnamefont{{Oxborrow}}},
  \bibinfo{author}{\bibfnamefont{S.}~\bibnamefont{{Pugla}}},
  \bibinfo{author}{\bibfnamefont{J.}~\bibnamefont{{Millo}}}, \bibnamefont{and}
  \bibinfo{author}{\bibfnamefont{P.}~\bibnamefont{{Gill}}},
  \bibinfo{journal}{Phys. Rev. A} \textbf{\bibinfo{volume}{77}},
  \bibinfo{pages}{033847} (\bibinfo{year}{2008}).

\bibitem[{\citenamefont{Siegman}(1986)}]{Siegman}
\bibinfo{author}{\bibfnamefont{A.}~\bibnamefont{Siegman}},
  \emph{\bibinfo{title}{Lasers}} (\bibinfo{publisher}{University Science
  Books}, \bibinfo{year}{1986}).

\bibitem[{\citenamefont{{Turnbull} et~al.}(1996)\citenamefont{{Turnbull},
  {Robertson}, {Smith}, {Allen}, and {Padgett}}}]{Turnbull96}
\bibinfo{author}{\bibfnamefont{G.~A.} \bibnamefont{{Turnbull}}},
  \bibinfo{author}{\bibfnamefont{D.~A.} \bibnamefont{{Robertson}}},
  \bibinfo{author}{\bibfnamefont{G.~M.} \bibnamefont{{Smith}}},
  \bibinfo{author}{\bibfnamefont{L.}~\bibnamefont{{Allen}}}, \bibnamefont{and}
  \bibinfo{author}{\bibfnamefont{M.~J.} \bibnamefont{{Padgett}}},
  \bibinfo{journal}{Optics Communications} \textbf{\bibinfo{volume}{127}},
  \bibinfo{pages}{183} (\bibinfo{year}{1996}).

\bibitem[{\citenamefont{{Courtial} and {Padgett}}(1999)}]{Courtial99}
\bibinfo{author}{\bibfnamefont{J.}~\bibnamefont{{Courtial}}} \bibnamefont{and}
  \bibinfo{author}{\bibfnamefont{M.~J.} \bibnamefont{{Padgett}}},
  \bibinfo{journal}{Optics Communications} \textbf{\bibinfo{volume}{159}},
  \bibinfo{pages}{13} (\bibinfo{year}{1999}).

\bibitem[{\citenamefont{{Kennedy} et~al.}(2002)\citenamefont{{Kennedy},
  {Szabo}, {Teslow}, {Porterfield}, and {Abraham}}}]{Kennedy02}
\bibinfo{author}{\bibfnamefont{S.~A.} \bibnamefont{{Kennedy}}},
  \bibinfo{author}{\bibfnamefont{M.~J.} \bibnamefont{{Szabo}}},
  \bibinfo{author}{\bibfnamefont{H.}~\bibnamefont{{Teslow}}},
  \bibinfo{author}{\bibfnamefont{J.~Z.} \bibnamefont{{Porterfield}}},
  \bibnamefont{and} \bibinfo{author}{\bibfnamefont{E.~R.}
  \bibnamefont{{Abraham}}}, \bibinfo{journal}{Phys. Rev. A}
  \textbf{\bibinfo{volume}{66}}, \bibinfo{pages}{043801}
  (\bibinfo{year}{2002}).

\bibitem[{\citenamefont{{Freise} and {Strain}}(2010)}]{freise10}
\bibinfo{author}{\bibfnamefont{A.}~\bibnamefont{{Freise}}} \bibnamefont{and}
  \bibinfo{author}{\bibfnamefont{K.}~\bibnamefont{{Strain}}},
  \bibinfo{journal}{Living Reviews in Relativity}
  \textbf{\bibinfo{volume}{13}}, \bibinfo{pages}{1} (\bibinfo{year}{2010}),
  \urlprefix\url{http://relativity.livingreviews.org/Articles/lrr-2010-1/}.

\bibitem[{\citenamefont{{Arlt} et~al.}(1998)\citenamefont{{Arlt}, {Dholakia},
  {Allen}, and {Padgett}}}]{Arlt98}
\bibinfo{author}{\bibfnamefont{J.}~\bibnamefont{{Arlt}}},
  \bibinfo{author}{\bibfnamefont{K.}~\bibnamefont{{Dholakia}}},
  \bibinfo{author}{\bibfnamefont{L.}~\bibnamefont{{Allen}}}, \bibnamefont{and}
  \bibinfo{author}{\bibfnamefont{M.~J.} \bibnamefont{{Padgett}}},
  \bibinfo{journal}{Journal of Modern Optics} \textbf{\bibinfo{volume}{45}},
  \bibinfo{pages}{1231} (\bibinfo{year}{1998}).

\bibitem[{\citenamefont{{Matsumoto} et~al.}(2008)\citenamefont{{Matsumoto},
  {Ando}, {Inoue}, {Ohtake}, {Fukuchi}, and {Hara}}}]{Matsumoto08}
\bibinfo{author}{\bibfnamefont{N.}~\bibnamefont{{Matsumoto}}},
  \bibinfo{author}{\bibfnamefont{T.}~\bibnamefont{{Ando}}},
  \bibinfo{author}{\bibfnamefont{T.}~\bibnamefont{{Inoue}}},
  \bibinfo{author}{\bibfnamefont{Y.}~\bibnamefont{{Ohtake}}},
  \bibinfo{author}{\bibfnamefont{N.}~\bibnamefont{{Fukuchi}}},
  \bibnamefont{and} \bibinfo{author}{\bibfnamefont{T.}~\bibnamefont{{Hara}}},
  \bibinfo{journal}{Journal of the Optical Society of America A}
  \textbf{\bibinfo{volume}{25}}, \bibinfo{pages}{1642} (\bibinfo{year}{2008}).

\bibitem[{\citenamefont{{R{\"u}diger} et~al.}(1981)\citenamefont{{R{\"u}diger},
  {Schilling}, {Schnupp}, {Winkler}, {Billing}, and
  {Maischberger}}}]{Ruediger81}
\bibinfo{author}{\bibfnamefont{A.}~\bibnamefont{{R{\"u}diger}}},
  \bibinfo{author}{\bibfnamefont{R.}~\bibnamefont{{Schilling}}},
  \bibinfo{author}{\bibfnamefont{L.}~\bibnamefont{{Schnupp}}},
  \bibinfo{author}{\bibfnamefont{W.}~\bibnamefont{{Winkler}}},
  \bibinfo{author}{\bibfnamefont{H.}~\bibnamefont{{Billing}}},
  \bibnamefont{and}
  \bibinfo{author}{\bibfnamefont{K.}~\bibnamefont{{Maischberger}}},
  \bibinfo{journal}{Optica Acta} \textbf{\bibinfo{volume}{28}},
  \bibinfo{pages}{641} (\bibinfo{year}{1981}).
  
\bibitem[{\citenamefont{et~al.}(2004)}]{Abbott04}
\bibinfo{author}{\bibfnamefont{B. Abbott} \bibnamefont{et~al.}},
  \bibinfo{journal}{Nuclear Instruments and Methods in Physics Research Section
  A: Accelerators, Spectrometers, Detectors and Associated Equipment}
  \textbf{\bibinfo{volume}{517}}, \bibinfo{pages}{154 } (\bibinfo{year}{2004}),
  ISSN \bibinfo{issn}{0168-9002},
  \urlprefix\url{http://www.sciencedirect.com/science/article/B6TJM-4B5R97X-F/%
2/7a87918172dbb84942fb67b139b8bd28}.

\bibitem[{\citenamefont{{Go{\ss}ler} et~al.}(2003)\citenamefont{{Go{\ss}ler},
  {Casey}, {Freise}, {Grant}, {Grote}, {Heinzel}, {Heurs}, {Husman},
  {K{\"o}tter}, {Leonhardt} et~al.}}]{gossler03}
\bibinfo{author}{\bibfnamefont{S.}~\bibnamefont{{Go{\ss}ler}}},
  \bibinfo{author}{\bibfnamefont{M.~M.} \bibnamefont{{Casey}}},
  \bibinfo{author}{\bibfnamefont{A.}~\bibnamefont{{Freise}}},
  \bibinfo{author}{\bibfnamefont{A.}~\bibnamefont{{Grant}}},
  \bibinfo{author}{\bibfnamefont{H.}~\bibnamefont{{Grote}}},
  \bibinfo{author}{\bibfnamefont{G.}~\bibnamefont{{Heinzel}}},
  \bibinfo{author}{\bibfnamefont{M.}~\bibnamefont{{Heurs}}},
  \bibinfo{author}{\bibfnamefont{M.~E.} \bibnamefont{{Husman}}},
  \bibinfo{author}{\bibfnamefont{K.}~\bibnamefont{{K{\"o}tter}}},
  \bibinfo{author}{\bibfnamefont{V.}~\bibnamefont{{Leonhardt}}},
  \bibnamefont{et~al.}, \bibinfo{journal}{Review of Scientific Instruments}
  \textbf{\bibinfo{volume}{74}}, \bibinfo{pages}{3787} (\bibinfo{year}{2003}).
  
\bibitem[{\citenamefont{{Genin} et~al.}(2010)\citenamefont{{Genin}, {Marque},
  {Swinkels}, and {Vajente}}}]{Genin10}
\bibinfo{author}{\bibfnamefont{A.}~\bibnamefont{{Genin}}},
  \bibinfo{author}{\bibfnamefont{J.}~\bibnamefont{{Marque}}},
  \bibinfo{author}{\bibfnamefont{B.}~\bibnamefont{{Swinkels}}},
  \bibnamefont{and}
  \bibinfo{author}{\bibfnamefont{G.}~\bibnamefont{{Vajente}}},
  \bibinfo{type}{Tech. Rep.} \bibinfo{number}{VIR-0232A-10},
  \bibinfo{institution}{Virgo} (\bibinfo{year}{2010}).

\bibitem[{\citenamefont{LIGO Laboratory and the LIGO Scientific Collaboration}(2009)}]{AdLIGOrefdesign}
  \bibinfo{author}{\bibnamefont{{Advanced LIGO Team}}}, \bibinfo{type}{Tech.
  Rep.},\bibinfo{number}{LIGO-M060056-v1} \bibinfo{institution}{LSC} (\bibinfo{year}{2009}).

\bibitem[{ast()}]{astigm}
\emph{\bibinfo{title}{\rm paper in preparation}}.

\bibitem[{\citenamefont{Skettrup}(2005)}]{skettrup05}
\bibinfo{author}{\bibfnamefont{T.}~\bibnamefont{Skettrup}},
  \bibinfo{journal}{Journal of Optics A: Pure and Applied Optics}
  \textbf{\bibinfo{volume}{7}}, \bibinfo{pages}{645} (\bibinfo{year}{2005}),
  \urlprefix\url{http://stacks.iop.org/1464-4258/7/645}.


\bibitem[{\citenamefont{Black}(2000)}]{Black00}
\bibinfo{author}{\bibfnamefont{E.~D.} \bibnamefont{Black}},
  \bibinfo{journal}{Am. J. Phys.} \textbf{\bibinfo{volume}{69}},
  \bibinfo{pages}{79} (\bibinfo{year}{2000}).

\bibitem[{\citenamefont{Freise et~al.}(2004)\citenamefont{Freise, Heinzel,
  L\"{u}ck, Schilling, Willke, and Danzmann}}]{Finesse}
\bibinfo{author}{\bibfnamefont{A.}~\bibnamefont{Freise}},
  \bibinfo{author}{\bibfnamefont{G.}~\bibnamefont{Heinzel}},
  \bibinfo{author}{\bibfnamefont{H.}~\bibnamefont{L\"{u}ck}},
  \bibinfo{author}{\bibfnamefont{R.}~\bibnamefont{Schilling}},
  \bibinfo{author}{\bibfnamefont{B.}~\bibnamefont{Willke}}, \bibnamefont{and}
  \bibinfo{author}{\bibfnamefont{K.}~\bibnamefont{Danzmann}},
  \bibinfo{journal}{Classical and Quantum Gravity}
  \textbf{\bibinfo{volume}{21}}, \bibinfo{pages}{S1067} (\bibinfo{year}{2004}),
  \bibinfo{note}{software available at:
  \href{http://www.gwoptics.org/finessef}{http://www.gwoptics.org/finesse}}.

\bibitem[{\citenamefont{Kwee et~al.}(2007)\citenamefont{Kwee, Seifert, Willke,
  and Danzmann}}]{KSWD07}
\bibinfo{author}{\bibfnamefont{P.}~\bibnamefont{Kwee}},
  \bibinfo{author}{\bibfnamefont{F.}~\bibnamefont{Seifert}},
  \bibinfo{author}{\bibfnamefont{B.}~\bibnamefont{Willke}}, \bibnamefont{and}
  \bibinfo{author}{\bibfnamefont{K.}~\bibnamefont{Danzmann}},
  \bibinfo{journal}{Review of Scientific Instruments}
  \textbf{\bibinfo{volume}{78}}, \bibinfo{pages}{073103}
  (\bibinfo{year}{2007}).

\bibitem[{\citenamefont{Willke et~al.}(1998)\citenamefont{Willke, Uehara,
  Gustafson, Byer, King, Seel, and R.~L.~Savage}}]{WUGBKSS98}
\bibinfo{author}{\bibfnamefont{B.}~\bibnamefont{Willke}},
  \bibinfo{author}{\bibfnamefont{N.}~\bibnamefont{Uehara}},
  \bibinfo{author}{\bibfnamefont{E.~K.} \bibnamefont{Gustafson}},
  \bibinfo{author}{\bibfnamefont{R.~L.} \bibnamefont{Byer}},
  \bibinfo{author}{\bibfnamefont{P.~J.} \bibnamefont{King}},
  \bibinfo{author}{\bibfnamefont{S.~U.} \bibnamefont{Seel}}, \bibnamefont{and}
  \bibinfo{author}{\bibfnamefont{J.}~\bibnamefont{R.~L.~Savage}},
  \bibinfo{journal}{Opt. Lett.} \textbf{\bibinfo{volume}{23}},
  \bibinfo{pages}{1704} (\bibinfo{year}{1998}).

\end{thebibliography}
\def\gobble, .{\ignorespaces}

\end{document}